\documentclass{edm_article}

\usepackage{multirow}
\usepackage{array}
\newcolumntype{P}[1]{>{\centering\arraybackslash}p{#1}}

\begin{document}

\title{Multimodal Assessment of Classroom Discourse Quality: A Text-Centered Attention-Based Multi-Task Learning Approach}


%
%
%
%

\numberofauthors{1} 
%
\author{
%
%
\alignauthor Ruikun~Hou\textsuperscript{\normalsize 1,2}, Babette B\"uhler\textsuperscript{\normalsize 1}, Tim F\"utterer\textsuperscript{\normalsize 2}, Efe~Bozkir\textsuperscript{\normalsize 1}, Peter~Gerjets\textsuperscript{\normalsize 3}, Ulrich~Trautwein\textsuperscript{\normalsize 2}, and Enkelejda~Kasneci\textsuperscript{\normalsize 1}\\
       \affaddr{\textsuperscript{1}Technical University of Munich, \textsuperscript{2}University of Tübingen, \textsuperscript{3}Leibniz-Institut für Wissensmedien}\\
       \email{\{ruikun.hou, babette.buehler, efe.bozkir, enkelejda.kasneci\}@tum.de, \{tim.fuetterer, ulrich.trautwein\}@uni-tuebingen.de, p.gerjets@iwm-tuebingen.de}
}

\maketitle

\begin{abstract}
Classroom discourse is an essential vehicle through which teaching and learning take place. Assessing different characteristics of discursive practices and linking them to student learning achievement enhances the understanding of teaching quality. Traditional assessments rely on manual coding of classroom observation protocols, which is time-consuming and costly. Despite many studies utilizing AI techniques to analyze classroom discourse at the utterance level, investigations into the evaluation of discursive practices throughout an entire lesson segment remain limited. Existing discourse assessment approaches primarily depend on transcript-based analyses, neglecting non-verbal modalities crucial for comprehensive evaluation.
To address this gap, our study proposes a novel text-centered multimodal fusion architecture to assess the quality of three discourse components grounded in the Global Teaching InSights (GTI) observation protocol: \textit{Nature of Discourse}, \textit{Questioning}, and \textit{Explanations}. First, we employ attention mechanisms to capture inter- and intra-modal interactions from transcript, audio, and video streams. Second, a multi-task learning approach is adopted to jointly predict the quality scores of the three components. Third, we formulate the task as an ordinal classification problem to account for rating level order. The effectiveness of these designed elements is demonstrated through an ablation study on the GTI Germany dataset containing 92 videotaped math lessons. Our results highlight the dominant role of text modality in approaching this task. Integrating acoustic features enhances the model's consistency with human ratings, achieving an overall Quadratic Weighted Kappa score of 0.384, comparable to human inter-rater reliability (0.326). Furthermore, correlation analyses between predicted ratings and student outcomes (i.e., test scores, interest, and self-efficacy) reveal partial alignment with those derived from human ratings. Our study lays the groundwork for the future development of automated discourse quality assessment to support teacher professional development through timely feedback on multidimensional discourse practices. 
\end{abstract}


\section{Introduction}

Classroom discourse serves as the medium through which teaching and learning take place \cite{cazden1988classroom}. 
Previous research has consistently shown that the quality of classroom discourse is positively associated with student learning \cite{mercer2014} and that teachers' skills in facilitating productive discourse influence students' motivation, autonomy, and interest \cite{kiemer2015effects}.
Hence, a thorough understanding of discourse dynamics guides teachers toward a more effective communicative practice \cite{walsh2006investigating}, holding the potential to inform and enhance teacher professional development programs. 
Given its importance, discourse quality represents a critical domain in structured classroom observation protocols, which are widely used by educational researchers to systematically evaluate teaching practices. These protocols such as CLASS (Classroom Assessment Scoring System \cite{pianta2008classroom}) and PLATO (Protocol for Language Arts Teaching Observations \cite{grossman2013measure}) assess multiple aspects of teaching quality, including classroom management, social-emotional support, and instructional discourse.
In this work, we adopt the observation protocol developed by the Global Teaching InSights (GTI) study \cite{gti2020}, a large-scale international classroom video study. The GTI framework operationalizes discourse quality through three specific components: \textit{Nature of Discourse}, which examines whether classroom interactions are teacher-directed or student-centered; \textit{Questioning}, which evaluates the cognitive demand of teacher questions, ranging from simple recall to higher-order thinking; and \textit{Explanations}, which assess the depth and clarity of mathematical descriptions provided by both teachers and students during instruction.
While these components do not encompass all aspects of discourse quality, they capture essential features of classroom discourse, aligning with established educational research that emphasizes the importance of student participation in discussion \cite{alexander2008towards,franke2007mathematics}, cognitive activation through strategic teacher questioning \cite{chin2007teacher,redfield1981meta}, and detailed mathematical explanations for student understanding \cite{hill2005effects,webb2014engaging}.

Traditionally, the implementation of structured protocols relies on human observers who watch videotaped lessons and assign scores for multiple dimensions according to associated rubrics.
Whereas these protocols have demonstrated validity in various large-scale classroom studies \cite{kane2013have,gti2020}, the manual coding process is accompanied by several challenges that limit their widespread application. 
First, the holistic assessment of teaching quality throughout an entire lesson segment often involves high-level inference about complex classroom interactions, which can lead to subjective evaluations and limited inter-rater reliability \cite{whitehill2023automated,demszky2023can}. Second, recent studies have demonstrated that the patterns frequently identified in observation-based teaching quality assessments do not always correspond with results from alternative data sources like student self-reports \cite{white2023}. Third, the manual nature of these evaluations, along with the need for extensive rater training, makes them a resource-intensive endeavor in terms of time and cost.

To address these challenges, recent research has increasingly turned to artificial intelligence (AI) techniques to automatically codify classroom observation protocols across various dimensions  \cite{ramakrishnan2021toward,wang2023chatgpt,hou2024automated,whitehill2023automated,tran2024analyzing}.
These studies aim to provide teachers with reliable and timely feedback on their teaching practices, with particular potential to foster teacher professional growth. 
With respect to automated assessment of instructional discourse, most existing studies have focused on analyzing transcripts employing Large Language Models (LLMs) \cite{wang2023chatgpt,whitehill2023automated,tran2024analyzing}. 
However, classroom discourse involves not only linguistic interactions but also non-verbal elements like gestures and silence \cite{Tsui2008}, which remain unexplored in these transcript-based approaches.
Therefore, we propose a novel multimodal fusion architecture for classroom discourse assessment grounded in the GTI observation framework, offering a more comprehensive understanding of discursive practices.

Automatically assessing discourse quality based on observation protocols poses several computational challenges.
First, measures for teaching effectiveness are typically associated with observable behavioral indicators involving verbal, acoustic, and visual cues, necessitating the use of techniques capable of effectively processing and integrating multiple modalities.
Second, since the data unit is long sequential signals that often span several minutes, algorithms should be able to capture long-range temporal dynamics.
Third, as the discourse domain consists of multiple dimensions measuring distinct aspects of communicative practices, computational efficiency is an essential factor beyond predictive accuracy, to ensure the practical applicability of these methods.
To tackle these challenges, we propose a novel architecture to assess the three discourse components—\textit{Nature of Discourse}, \textit{Questioning}, and \textit{Explanations}—by integrating transcript, audio, and video streams through attention mechanisms \cite{attention}. Attention mechanisms, the foundation of advanced transformer models like ChatGPT, have proven exceptional performance in processing sequential data. 
The textual stream is treated as primary, since the rubrics for our focal components, especially \textit{Questioning} and \textit{Explanations}, are content-oriented. We retain the acoustic and visual channels, as they may capture complementary cues that shape how content is experienced by classroom participants.
Additionally, we adopt a multi-task learning approach to simultaneously predict the three components in a single pass. Since these components are interconnected, multi-task learning enables the model to capture their interdependencies while improving computational efficiency by avoiding the need to train separate models for each component.

Our contributions are summarized as follows: (1) We introduce a novel text-centered multimodal multi-task learning architecture that jointly predicts the quality of classroom discourse across three dimensions. We apply attention mechanisms to capture inter- and intra-modal interactions in classroom observation settings, which, to our knowledge, is the first application of such techniques in this context.
(2) Relying on a dataset involving 92 videotaped authentic lessons, we demonstrate the applicability of the proposed approach in real-world classroom environments and its efficacy by comparing it to human inter-rater reliability.  
(3) Through an extensive ablation study, we assess the contributions of diverse modality configurations and the effectiveness of different elements used in our architecture.
(4) We further investigate how the human-rated and model-predicted quality of classroom discourse is related to student learning and non-cognitive outcomes.

\section{Related Work}
In this section, we review related research in two areas: the automated classroom utterance analysis and the holistic assessment of teaching effectiveness using classroom observation protocols.

\subsection{Automated Classroom Discourse Analysis}
To characterize discursive strategies that can promote effective teaching, educational researchers have developed various theoretical frameworks, such as \textit{dialogic teaching}, which emphasizes the importance of teachers encouraging students to actively engage in discussions about learning content \cite{alexander2008towards} and \textit{accountable talk} in which students' academic language development is supported by providing students with key phrases they can during learning like ``explaining'', ``agreeing'', ``disagreeing'', or ``justifying'' \cite{o2015scaling}.
Grounded on these theoretical insights, many computational researchers have leveraged Natural Language Processing (NLP) techniques to automatically detect pedagogical strategies from individual utterances or turns within classroom dialogue \cite{hunkins2022beautiful,suresh2022talkmoves,wang2023teacher,demszky2021measuring,alic2022computationally,kupor2023measuring}. 
Suresh et al. \cite{suresh2022talkmoves} introduced the TalkMoves dataset in which ten dialog acts were annotated based on accountable talk theory \cite{o2015scaling}, e.g., whether a teacher talk promotes students to explain their ideas. They further employed language models such as BERT \cite{devlin2019bert} to classify each student-teacher or student-student utterance pair into one of these acts.
Recent works have expanded the prediction of accountable talk moves by either zero-shot prompting \cite{wang2023teacher} or fine-tuning \cite{kupor2023measuring} of LLMs.
Other researchers have focused on measuring different dimensions of teacher utterances, such as uptake of student contributions \cite{demszky2021measuring}, questioning strategies \cite{alic2022computationally}, and supportive discourse \cite{hunkins2022beautiful}.

While these approaches effectively capture fine-grained patterns at the utterance level, they do not account for the broader discourse dynamics that unfold over extended lesson periods.
Metrics that focus on specific teaching behaviors, such as the frequency of teacher questions during class, often provide limited practical guidance for educators who seek to improve their instructional strategies. 
Higher-level feedback, which aligns better with real-world teacher coaching practices, might be more valuable in this context \cite{wang2023chatgpt}. 

\subsection{Automated Classroom Observation}
Compared to individual utterance analysis, research on the automated assessment of holistic teaching effectiveness scores derived from classroom observation protocols remains relatively underexplored. 
James et al. \cite{james2018inferring} initially attempted to infer classroom positive and negative climate according to the CLASS protocol by extracting facial expressions and speech features from video and audio recordings. 
They simplified the task to a binary classification problem and aggregated the extracted features over time, fed into non-temporal classifiers like Random Forest for prediction.
Ramakrishnan et al. \cite{ramakrishnan2021toward} advanced this research direction by predicting fine-grained classroom climate categories corresponding to the 7-point rating scale in CLASS. They utilized temporal models like Long Short-Term Memory (LSTM) to capture sequential patterns over 15-minute recordings, leading to better performance than feature aggregation methods. 
Focusing on the same domain of social-emotional support, Hou et al. \cite{hou2024automated} additionally took into account textual sentiment features alongside facial and speech emotions, training non-temporal regressors to estimate quality scores of encouragement and warmth in classrooms. 

In addition to these multimodal approaches, another strand of research relies on the zero-shot capabilities of LLMs to score transcripts for different teaching practice domains \cite{wang2023chatgpt,hou2024automated,whitehill2023automated,tran2024analyzing}.
While these studies have highlighted the potential of recent LLMs in codifying observation protocols, they present several limitations.
First, LLM-based methods mainly focused on transcript analysis, neglecting the full potential of multimodal data. 
Besides, supervised learning models tailored to specific domain tasks tend to outperform LLMs operating as zero-shot generalists \cite{hou2024automated}.
Moreover, using commercial LLMs like ChatGPT requires access to APIs, which involves cost and might raise privacy concerns. 
Open-source alternatives, such as Llama \cite{dubey2024llama}, not only demand computational resources for deployment but still lag behind commercial competitors in terms of zero-shot performance \cite{ateia2024can}. 
Therefore, this work explores a supervised learning method.

As introduced above, existing multimodal supervised approaches have concentrated on assessing various aspects of classroom climate, with no specific attention given to discourse quality.
These methods also have several limitations: 
First, transcripts involving rich conversational exchanges between teachers and students were not considered in previous studies \cite{james2018inferring,ramakrishnan2021toward}. 
Second, multimodal fusion was often achieved through basic feature concatenation \cite{hou2024automated} or by averaging predictions from independent unimodal pathways \cite{ramakrishnan2021toward}, limiting the modeling of detailed cross-modal interactions.
Third, temporal dynamics of classroom interactions were insufficiently captured due to the reliance on non-temporal models \cite{james2018inferring,hou2024automated}.
Finally, most studies framed the assessment as a standard classification task, overlooking the ordinal nature of rating scales \cite{james2018inferring,ramakrishnan2021toward}. 

To address the limitations, we introduce a novel architecture that leverages state-of-the-art pre-trained models to extract verbal, acoustic, and visual features from transcript, audio, and video data. We then adopt cross- and self-attention encoders to integrate multimodal representations while modeling temporal dependencies. Further, we implement a loss function designed explicitly for ordinal classification and propose simultaneously estimating multiple dimensions of discourse quality via multi-task learning.
Finally, we examine the correlation between automated assessments and student outcomes, an aspect unexplored in previous research.

\section{Method}
In this section, we present our methodology for automated classroom discourse assessment. We begin by describing the Global Teaching InSights (GTI) dataset used in this study, highlighting the three discourse components within its observation protocol. Next, we provide a brief overview of our task formulation and present the key elements of our proposed architecture in detail, including pre-trained models for unimodal feature extraction, attention-based multimodal fusion, and multi-task learning with ordinal classification.
\subsection{Dataset}
\label{sect:dataset}
\subsubsection{GTI Dataset}
GTI is an international classroom observation study involving eight countries \cite{gti2020}, initially referred to as the Teaching and Learning International Survey (TALIS). This large-scale study aims to investigate the quality of instructional practices and their impact on student learning and non-cognitive outcomes. Focusing on the teaching and learning of a specific mathematics topic (i.e., quadratic equations), the study systematically collected a wide range of data from real-world classrooms, including videotaped lessons, instructional artifacts such as worksheets, student tests, as well as teacher and student questionnaires. 

In this work, we utilized the German subset from GTI, comprising 100 videotaped math lessons from 50 recruited teachers (46\% female, average age: 43 years) and over 1,140 students (53\% female, average age: 15 years) across 38 schools. Due to data protection restrictions, only 92 recordings were accessible for this analysis. 
The recordings were conducted with a stationary camera operating at 25 frames per second (FPS) to capture as many classroom participants as possible. 
Additionally, GTI provided human-transcribed lesson transcripts, annotated by timestamps and anonymized speaker IDs (e.g., ``L'' for the teacher and ``S01'' for a student) for each conversation turn.

\subsubsection{Discourse Components in GTI Observation Protocol}
\label{subsubsect:discourse}
Among the various coding instruments developed by GTI to quantify teaching quality based on diverse data sources, we employed the video observation protocol \cite{gti2018codes} in this work. 
This structured protocol consists of six domains that offer a general overview of the relevant teaching practices, including \textit{Classroom Management}, \textit{Social-Emotional Support}, \textit{Discourse}, \textit{Quality of Subject Matter}, \textit{Student Cognitive Engagement}, and \textit{Assessment of and Responses to Student Understanding}. 
Each domain is divided into three components, capturing the quality of specific teaching constructs that involve coarser-grained patterns of behavior over extended lesson periods. 
As specified in the coding protocol, videos are segmented into 16-minute intervals, with each segment scored on a scale of 1-4 by extensively trained observers for 18 components. 
A 16-minute window offers an appropriate balance by being long enough to observe extended instructional patterns necessary for rating coarse-grained components, yet short enough to avoid excessive cognitive load on human observers and support reliable human coding. This segmentation also aligns with other classroom observation frameworks such as CLASS, which uses 15-minute windows.
Higher rating scores indicate higher levels of teaching effectiveness in the respective component. 

\begin{table*}
    \caption{Characteristics of high-quality teaching practices in the Discourse domain \cite{gti2018codes}.}
    \label{tab:components}
    \centering
    \renewcommand{\arraystretch}{1.2}
    \begin{tabular}{ll}
        \hline
        \textbf{Component} & \textbf{Characteristics of respective high-quality teaching practices} \\
        \hline
        Nature of Discourse & More student-centered discourse with frequent detailed contributions \\ \hline
        Questioning & Emphasis on questions that request students analyze, synthesize, justify, or conjecture \\ \hline
        Explanations & Explanations focus on lengthy/deeper features of the mathematics. \\
        \hline
    \end{tabular}
\end{table*}

\begin{figure*}[h]
    \centering
    \includegraphics[width=0.9\textwidth]{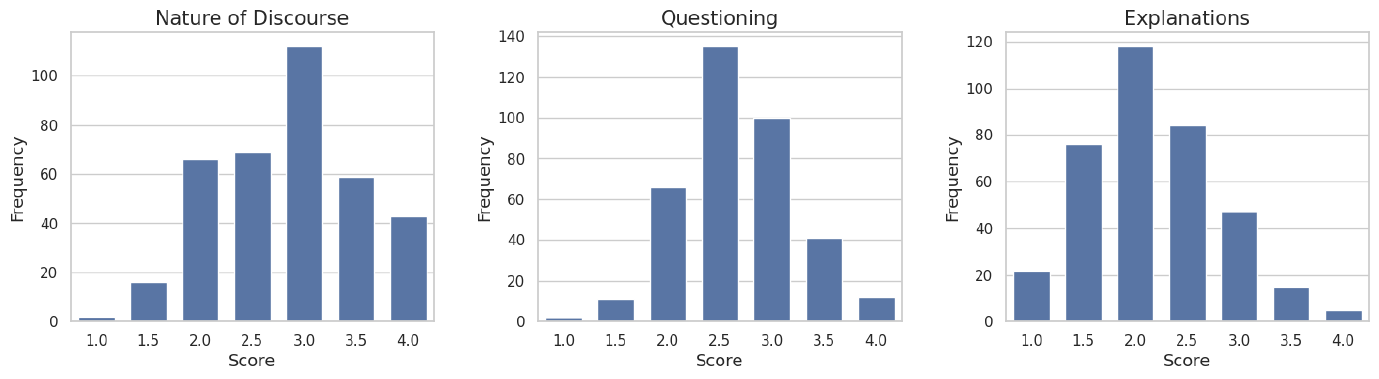}
    \caption{Distributions of average double-coded human ratings for three discourse components.}
    \label{fig:human_rating_distribution}
\end{figure*}

As instructional discourse directly shapes student's learning experiences \cite{cazden1988classroom} and no prior computational research has explored its assessment using multimodal data streams, our study was dedicated to the \textit{Discourse} domain within the GTI protocol, which includes three components: \textit{Nature of Discourse}, \textit{Questioning}, and \textit{Explanations}. 
Overall, classroom discourse refers to any form of communication between and among the teacher and students \cite{gti2018trainingnotes}. These three components are characterized in the GTI protocol as follows:
\begin{enumerate}
    \item \textit{Nature of Discourse} distinguishes teacher-directed and more student-centered interactions. Teacher-directed discourse is characterized by a monological lecture style or by the teacher controlling the communication flow through questions and answers. In contrast, student-centered discourse involves students actively contributing detailed insights on the subject matter, with the teacher taking a less dominant role in shaping the conversation.
    \item \textit{Questioning} evaluates the types of questions the teacher poses, focusing on the level of cognitive demand placed on students. The scale spans from basic recall or yes/no questions to more complex inquiries that require students to explain, classify, analyze, or synthesize information.
    \item The \textit{Explanations} component examines the extent to which teachers and students provide justifications or reasoning behind mathematical ideas and procedures. It assesses the depth of these explanations, from superficial or brief descriptions to more detailed and thorough explorations of the subject matter.
\end{enumerate}
Table~\ref{tab:components} presents a summary of the key attributes that distinguish high-quality teaching practices for each discourse component (refer to \cite{gti2018codes,gti2018trainingnotes} for more details regarding the coding rubrics, definitions, and behavioral examples of the three components).

\subsubsection{Preprocessing}

Following the GTI observation protocol, we preprocessed the data by splitting lesson recordings and transcripts into 16-minute segments. If the last segment of a recording was shorter than eight minutes, it was combined with the preceding segment. This process yielded 367 segments, forming the dataset for developing and evaluating our automated estimation methods. 
In Germany, 14 raters participated in the coding process, with two raters randomly assigned to independently evaluate each lesson.
The average score across both raters was calculated for each segment, serving as the ground truth. 
To maintain data granularity and prevent information loss, the mean scores were not rounded to integers, resulting in seven distinct categorical ratings used in the subsequent analysis.
The distribution of these averaged ratings for each discourse component is depicted in Figure~\ref{fig:human_rating_distribution}.

\subsection{Problem Formulation}
Given the classroom dataset $\boldsymbol{D} = \left\{(\boldsymbol{S}_i, \hat{\boldsymbol{y}}_i) | i = 1, ..., N\right\}$, each of the $N = 367$ lesson segments $\boldsymbol{S}_i$ contains three types of unimodal raw signals: text $\boldsymbol{S}_i^t$, audio $\boldsymbol{S}_i^a$, and video $\boldsymbol{S}_i^v$. 
The ground truth for $\boldsymbol{S}_i$ is represented by $\hat{\boldsymbol{y}}_i = \left\{\hat{y}_i^n, \hat{y}_i^q, \hat{y}_i^e\right\}$, where $\hat{y}_i^c \in \boldsymbol{L}$, for $c \in \left\{n, q, e\right\}$, 
corresponds to the averaged double-rated scores for the three discourse components: \textit{Nature of Discourse}, \textit{Questioning}, and \textit{Explanations}, respectively. The label set $\boldsymbol{L} = \left\{1, 1.5, ..., 4\right\}$ consists of the seven categorical ratings.
This study aims to develop a supervised model that can accurately assess each discourse component $\hat{y}_i^c$ for any given segment $\boldsymbol{S}_i$, using the three available unimodal signals as input. 

\subsection{Multimodal Multi-Task Learning}

\begin{figure*}[h]
    \centering
    \includegraphics[width=\textwidth]{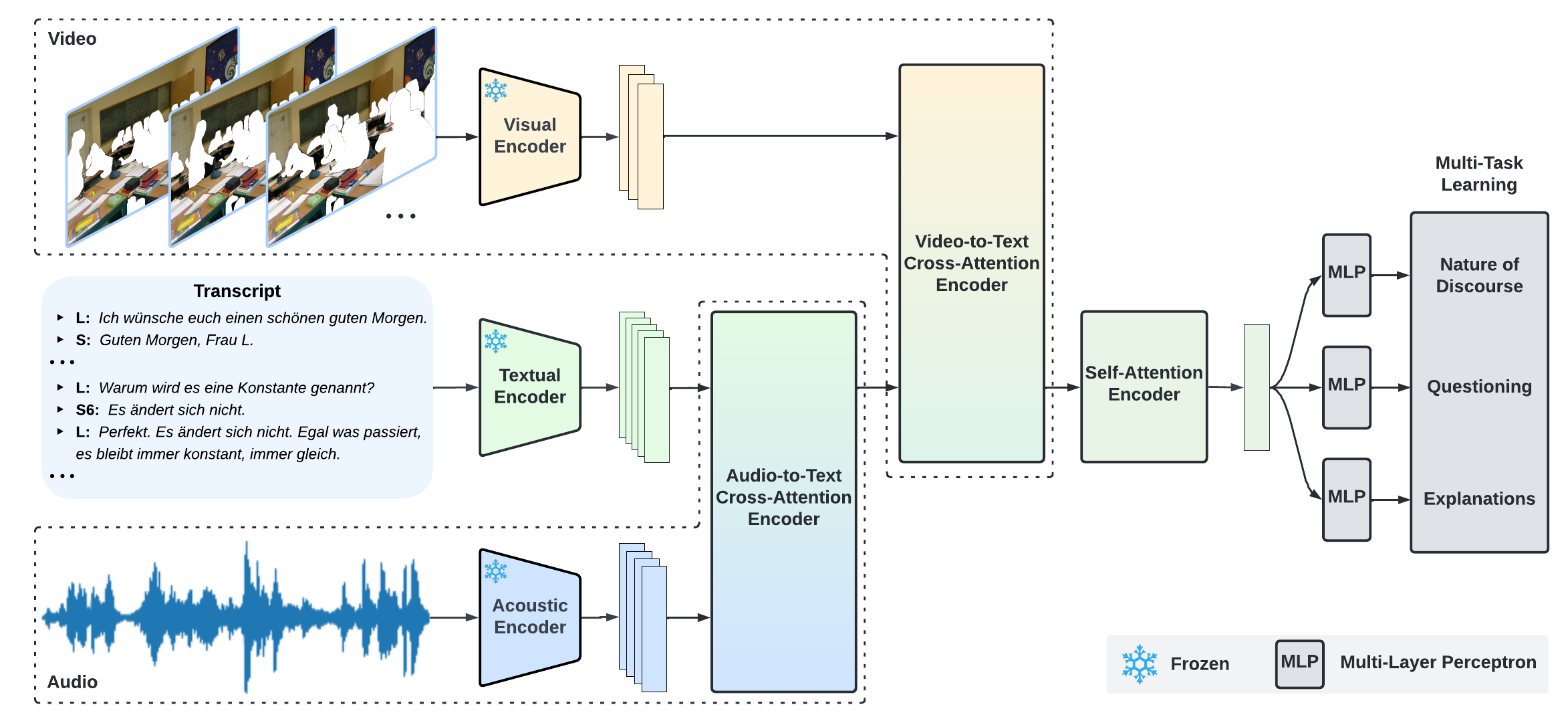}
    \caption{Text-centered multimodal attention-based architecture for classroom discourse assessment through multi-task learning.}
    \label{fig:architecture}
\end{figure*}

As illustrated in Figure~\ref{fig:architecture}, we propose a multimodal multi-task learning architecture to jointly estimate teaching effectiveness scores across three components of authentic classroom discourse in a single pass. 
In the initial step, we adopted pre-trained models to process the unimodal raw signals at an early stage. Considering the limited size of our dataset, their parameters were frozen during training to avoid overfitting.
Afterward, a series of attention-based intra- and inter-modal encoders were employed to extract highly representative embeddings guided by the textual modality. Lastly, three Multi-Layer Perceptrons (MLPs) were stacked on top to estimate scores for each discourse component. 
An ordinal classification loss function was implemented to address the ordering nature of the categorical ratings.
We present the detailed descriptions of each module in the subsequent subsections.

\subsubsection{Textual Encoder}
\label{subsubsec:text_encoder}
With the rapid development of NLP techniques, an increasing number of text embedding models have been introduced \cite{muennighoff2022mteb}.
To process our transcripts in German, we leveraged a state-of-the-art pre-trained embedding model, JinaBERT-Base-de\footnote{https://huggingface.co/jinaai/jina-embeddings-v2-base-de} \cite{mohr2024multi}, as the textural encoder. The model, built on the architecture of Bidirectional Encoder Representations from Transformers (BERT) \cite{devlin2019bert}, can handle bilingual input in both German and English. It supports input sequences of up to 8,192 tokens, making it suitable for our 16-minute transcripts, which often contain over 4,000 words. 

In particular, we applied the model to $\boldsymbol{S}_i^t$ to capture contextual information from the entire transcript, resulting in a 768-dimensional embedding for each word. Given the long transcript length, the direct use of these word embeddings in subsequent processing steps would yield a significant computational load. Thus, following the aggregation approach suggested by \cite{mohr2024multi}, we calculated utterance-level representations $\boldsymbol{X}_i^t \in \mathbb{R}^{L_i^t \times 768}$ by averaging the embeddings of all words within each utterance. Here, $L_i^t$ denotes the number of utterances in $\boldsymbol{S}_i^t$.

\subsubsection{Acoustic Encoder}
There are two primary approaches to extracting acoustic representations in the literature, i.e., low-level descriptors (e.g., Mel-Frequency Cepstral Coefficients (MFCCs)) and deep embeddings generated by pre-trained models (e.g., wav2vec 2.0 \cite{baevski2020wav2vec}).
Recent studies have shown that latent representations produced by deep networks often achieve superior performance over spectral features in various speech-relevant downstream tasks like emotion recognition \cite{wu2024multimodal,pepino21interspeech}. 
Besides, previous research had observed a noticeable performance drop when models' training and testing phases were conducted in different languages \cite{tamulevivcius2020study}. 
Considering these challenges, we employed a pre-trained model, XLSR-German\footnote{https://huggingface.co/facebook/wav2vec2-large-xlsr-53-german} proposed by Meta AI, to process raw audio signals. 
This model is designed for automatic speech recognition (ASR) tasks by fine-tuning XLSR \cite{conneau2020unsupervised} in German data. XLSR, a cross-lingual extension of wav2vec 2.0, is a transformer-based \cite{attention} model trained on enormous unlabeled speech data in 53 languages. 

Instead of using XLSR-German for ASR, we extracted the hidden states from the layer preceding the final inference layer. This approach transformed an auditory signal into a sequence of 1024-dimensional embeddings.
To ensure computational efficiency for long-duration audio processing, we divided each audio segment $\boldsymbol{S}_i^a$ into non-overlapping chunks, each lasting 10 seconds with a sampling rate of 16~kHz. This chunk duration was selected experimentally.
We then fed each chunk into XLSR-German and computed the average of the resulting embedding sequence. As a result, we obtained chunk-wise representations $\boldsymbol{X}_i^a \in \mathbb{R}^{L_i^a \times 1024}$, where $L_i^a$ represents the number of chunks in $\boldsymbol{S}_i^a$.

\subsubsection{Visual Encoder}
Previous research often focused on analyzing individual faces within a video frame and then aggregating the results to generate a frame representation, proving effective in affective computing applications \cite{shi2023multiemo,james2018inferring}. 
Nevertheless, this approach neglects scene-aware information, which may be significant for modeling multi-party interactions \cite{ramakrishnan2021toward}.
Additionally, individual face analysis in real-world classroom environments can suffer from frequent occlusions as well as low spatial resolution, especially when faces are far from the camera \cite{goldberg2021attentive,sumer2021multimodal,buhler2023automated}.
Therefore, we extracted semantically rich representations directly from a whole frame in this study.
To this end, we employed Contrastive Language-Image Pre-Training (CLIP) \cite{radford2021learning}, a dual-encoder model developed by OpenAI. Based on a large dataset of image-text pairs, CLIP attempts to simultaneously train textual and visual encoders to construct a shared embedding space. Since CLIP's textual encoder is designed for relatively short image captions, which is not suitable for our longer transcript data, we sorely utilized its visual encoder to process video frames. Specifically, we used the \textit{clip-vit-large-patch14-336}\footnote{https://huggingface.co/openai/clip-vit-large-patch14-336} model, which is widely applied in recent research on Multimodal Large Language Models (MLLMs) \cite{liu2024llava}.

The recording rate of 25 FPS resulted in a high degree of similarity between adjacent frames. To reduce computational redundancy, we uniformly sampled $\boldsymbol{S}_i^v$ at 1 FPS. The CLIP visual encoder was applied to each frame to extract a 768-dimensional embedding. We used a 10-second sliding window without overlapping to align with the auditory modality and averaged the frame embeddings within each window. This resulted in chunk-wise visual representations $\boldsymbol{X}_i^v \in \mathbb{R}^{L_i^v \times 768}$, where $L_i^v = L_i^a$ is the number of chunks in $\boldsymbol{S}_i^v$.

\subsubsection{Text-Centered Multimodal Fusion}
\label{subsubsec:fusion}
In line with the discourse components' definitions, which are closely relevant to verbal cues (see Sect.~\ref{subsubsect:discourse}), our pilot experiments revealed that models relying solely on transcripts outperformed those based on vision or audio. This highlights the primary significance of the textual modality for this task. 
However, the other two modalities can still capture complementary information that enhances the comprehension of classroom discourse.
For example, audio offers paralinguistic features such as pauses and speech speed, where longer wait time after questions and deliberate speech may indicate a classroom environment conducive to deeper thinking. 
Likewise, video data provides visual insights into teacher-student interaction patterns and body language like hand-raising, which reflects the level of student involvement and can potentially inform the assessment of \textit{Nature of Discourse}.
Considering these factors, we designed a text-centered multimodal fusion module utilizing the multi-head attention mechanism \cite{attention}. 

Inspired by BERT's \cite{devlin2019bert} use of a special classification (CLS) token for downstream tasks, we first prepended a learnable CLS embedding to the beginning of each unimodal representation $\boldsymbol{X}_i^{\left\{t,a,v\right\}}$.
Afterward, two cross-attention encoders were employed in sequence to integrate the audio $\boldsymbol{X}_i^a$ and video $\boldsymbol{X}_i^v$ representations into the textual embeddings $\boldsymbol{X}_i^t$, as depicted in Figure~\ref{fig:architecture}. In both encoders, the textual modality was consistently used as the Query (Q), while the corresponding auditory or visual modality served as the Key (K) and Value (V). 
This setup guides the model in focusing on the most relevant aspects of audio and visual data that align contextually with the transcript content, complementing textual understanding through inter-modal interactions. 
Followed by multimodal fusion, a self-attention encoder was applied to further learn intra-modal dependencies within the integrated textual embeddings, resulting in a refined representation $\boldsymbol{X}_i \in \mathbb{R}^{(L_i^t+1) \times 768}$. 
The self-attention mechanism allows the model to capture temporal dynamics and subtle patterns across the sequence of utterances. We note that all attention layers employed 12 heads.

To deepen the network, we sequentially stacked $M$ aforementioned fusion modules (i.e., one module comprises two cross-modal encoders and one self-attention encoder), with the output of each module $\boldsymbol{X}_i$ serving as the input to the next, forming the updated textural representation $\boldsymbol{X}_i^t$.
After passing through $M$ modules, we extracted the feature vector $\boldsymbol{x}_i \in \mathbb{R}^{768}$ from the first time step (i.e., corresponding to the CLS token) as the global representation of the entire sequence $\boldsymbol{X}_i$. This final vector integrated multimodal and temporal information for downstream tasks.

\subsubsection{Multi-Task Learning}
Although the three components measure distinct aspects of classroom discourse, they are likely interconnected. For instance, well-crafted questions that stimulate cognitive reasoning can enhance the depth of explanations and encourage students to make more detailed contributions.
Therefore, we employed a multi-task learning approach, with each task corresponding to assessing a specific component. This method offers several benefits: 
First, it accounts for the interdependencies between components by learning a shared representation that captures their common characteristics, thereby fostering the understanding of each component and improving the model's generalization abilities.
Second, by jointly balancing multiple objectives, this approach mitigates the risk of overfitting to any single task, potentially leading to better overall performance.
Third, multi-task learning avoids the need to train separate models for each component, reducing computational overhead. 

Specifically, we passed the final representation $\boldsymbol{x}_i$ through a set of component-specific MLPs to produce predictions $\boldsymbol{y}_i^c$, for $c \in \left\{n, q, e\right\}$. 
Each MLP consists of three fully connected layers with 512 hidden neurons and ReLU activation functions, followed by a Softmax layer to compute a probability distribution over the rating categories, i.e., $\boldsymbol{y}_i^c = \{p_{ij}^c|j=1, ..., K\}$ where $K=7$ is the number of rating categories and $p_{ij}^c$ represents the predicted probability for the $j$-th class of the $i$-th sample at component $c$. 
To tackle the ordinal classification problem, where categorical labels exhibit a natural order, we leveraged a novel loss function called Ordinal Log-Loss (OLL) \cite{castagnos2022simple}, defined as 
\begin{equation}
    L_{\text{OLL}}^c = - \frac{1}{N_t} \sum_{i=1}^{N_t} \left(w^c(\hat{y}_i^c) \sum_{j=1}^{K} \log(1 - p_{ij}^c) \left| \hat{y}_i^c - j \right|\right),
    \label{equ:oll}
\end{equation}
where $N_t$ is the number of training samples and $\left| \hat{y}_i^c - j \right|$ calculates the L1-norm distance between the true label $\hat{y}_i^c$ and the $j$-th class. 
To address the issue of imbalanced data (see Figure~\ref{fig:human_rating_distribution}), we applied class weights $w^c(\cdot)$ to each sample, computed as the inverse of class frequency within the training set. This enables the model to place more emphasis on minority-class samples during training. 
Compared to the widely used cross-entropy loss for standard classification, OLL considers the distance between categories, applying larger penalties for predictions that deviate significantly from the true category. 
For the multi-task learning setup, the total loss across all components is formulated as 
\begin{equation}
    L = \sum_{c \in \left\{n,q,e\right\}} \mu_c L_{\text{OLL}}^c,
\end{equation}
where $\mu_c$ is a weight parameter that balances the loss among tasks. We set $\mu_c=1$ to treat each component equally.

\section{Experiments}
\label{sec:experiments}
\subsection{Implementation}
\label{subsec:implementation}
We implemented a teacher-independent nested 5-fold cross-validation for training and evaluation. This approach grouped all the segments from the same teacher into a single fold, ensuring the model's generalizability to unseen teachers.
Besides, the cross-validation process allowed us to collect inferences for the entire dataset, as each segment was included in the test fold exactly once. We used these predictions for subsequent correlation analysis with student outcomes.
A fixed random seed was pre-set to ensure all models were evaluated on an identical fold split.
Further, we conducted grid search hyperparameter tuning within the inner cross-validation loop to find the optimal setting that maximizes average performance across three discourse components. Table~\ref{tab:hyperparameter} details the parameter space explored during tuning.
During model training, we set aside 20\% of training data as a validation set and implemented two strategies to prevent overfitting: learning rate reduction (halved after five epochs without validation loss improvement) and early stopping (triggered after 15 epochs without improvement).
We utilized the AdamW optimizer \cite{loshchilov2019adamw} with a weight decay factor of 0.01 to reduce overfitting during training.
Our experiments were carried out on an NVIDIA A100 GPU.

\begin{table}
    \caption{Grids for hyperparameter tuning.}
    \label{tab:hyperparameter}
    \centering
    \renewcommand{\arraystretch}{1.2}
    \begin{tabular}{ll}
        \hline
        \textbf{Parameter} & \textbf{Grid} \\
        \hline
        learning rate & 1e-4, 1e-5 \\
        batch size & 8, 16, 32 \\
        number of fusion modules ($M$) & 1, 2, 3, 4, 5 \\
        \hline
    \end{tabular}
\end{table}

\subsection{Evaluation Metrics}
We evaluated the model's performance regarding Quadratic Weighted Kappa (QWK) defined as
\begin{equation}
  \text{QWK} = 1 - \frac{\sum_{i=1}^K \sum_{j=1}^K w_{ij}O_{ij}}{\sum_{i=1}^K \sum_{j=1}^K w_{ij}E_{ij}},
\end{equation}
where $O_{ij}$ and $E_{ij}$ denote the observed and expected count in the $i$th row and $j$th column of the confusion matrix, respectively, and $w_{ij} = (i-j)^2$ is the corresponding weight.
Beyond classification accuracy, QWK provides a more fine-grained evaluation by penalizing predictions farther from the ground truth through $w_{ij}$, making it a suitable evaluation metric for our classroom discourse scoring task.

Following recent studies \cite{whitehill2023automated,hou2024automated}, we compared the alignment between model predictions and manual ratings to the agreement among human observers. 
To measure human inter-rater reliability (IRR), we employed a leave-one-rater-out method, where for each rater, QWK was calculated between their ratings and those from the other raters assigned to the same set of segments. We reported the average QWK across all $R$ raters, along with standard error estimates defined as the standard deviation of the average divided by $\sqrt{R}$. 
Regarding model evaluation, we calculated the average QWK between predictions and ground-truth scores over the five test folds and also reported the corresponding standard error estimates to illustrate performance stability.

\section{Results}
In this section, we present the results of automated classroom discourse assessment. Unless otherwise specified, all models were trained and evaluated under the same conditions described in Sect.~\ref{subsec:implementation}.

\begin{table*}
    \caption{Results for performance comparison with human inter-rater reliability and across modalities (T: text, A: audio, V: video) using QWK. For humans, the average QWK over raters is reported. For models, QWK is averaged over five folds. Standard error estimates are provided in parentheses, and the highest model-achieved QWK within each column is in bold.}
    \label{tab:results}
    \centering
    \renewcommand{\arraystretch}{1.2}
    \begin{tabular}{p{3.4cm}p{2cm}p{2.4cm}p{2.4cm}p{2.4cm}p{2.4cm}}
        \hline
        \textbf{Method} & \textbf{Modality} & \textbf{Nature of} \newline \textbf{Discourse} & \textbf{Questioning} & \textbf{Explanations} & \textbf{Average} \\
        \hline
        Human raters &  & 0.507 (0.03) & 0.228 (0.04) & 0.241 (0.04) & 0.326 (0.03) \\
        \hline
         \multirow{5}{*}{\parbox{3.4cm}{Proposed model\\(attention, \\multi-task learning, \\ordinal classification)}} 
         & T & 0.364 (0.03) & \textbf{0.387} (0.02) & \textbf{0.317} (0.05) & 0.356 (0.03) \\
         & A & 0.465 (0.03) & 0.163 (0.05) & 0.153 (0.08) & 0.261 (0.04) \\
         & V & 0.333 (0.09) & 0.035 (0.04) & 0.047 (0.07) & 0.138 (0.04) \\
         & T+A & \textbf{0.486} (0.04) & 0.382 (0.02) & 0.284 (0.04) & \textbf{0.384} (0.01)\\
         & T+A+V & 0.444 (0.03) & 0.344 (0.04) & 0.303 (0.07) & 0.364 (0.03) \\ 
         \hline
    \end{tabular}
\end{table*}

\subsection{Model Performance across Modalities}
Table~\ref{tab:results} summarizes the results of our multimodal multi-task learning framework in comparison to human IRR. 
To explore the effect of distinct modality combinations on classroom discourse assessment, we experimented with both unimodal and multimodal-fused representations. 
For unimodal experiments, we employed only the self-attention encoder. For multimodal configurations, the corresponding module enclosed by dashes in Figure~\ref{fig:architecture} was selectively included or omitted depending on which modalities were being fused.
Additionally, we computed the overall performance by averaging the scores from the three components for each rater (or fold in the case of models) and then averaging these values across all raters (or folds). We discuss the results below in detail.

\subsubsection{Human Inter-Rater Reliability}
Human observers achieved a moderate QWK score of 0.507 for the \textit{Nature of Discourse}, notably higher than the scores from other discourse components. This suggests that identifying whether a lesson is teacher-directed involves a lower level of inference than the less straightforward tasks of assessing the quality of posed questions or the depth of mathematical explanations.
Overall, the average QWK across all components yielded 0.326, revealing the inherent challenge of maintaining high consistency, even among experienced raters, when applying classroom observation protocols.

\subsubsection{Unimodal Performance}
Among unimodal approaches, the highest overall score of 0.356 was observed in the textural pathway, surpassing human IRR. Its strength lies in the effective assessment of the higher-inference components (\textit{Questioning} and \textit{Explanations}). 
Despite a lower overall agreement (0.261), the audio modality excelled in assessing \textit{Nature of Discourse}, likely due to its ability to capture relevant auditory indicators, such as the presence of multiple speakers and variations in their speech duration. 
In contrast, the video modality underperformed across all components, with an average QWK score of 0.138, indicating the low dependency of the discourse assessment on visual cues.

\subsubsection{Multimodal Performance}
Given the effectiveness of the textual modality, we proposed the text-guided cross-modal fusion approach described in Sect.~\ref{subsubsec:fusion}. 
As shown in Table~\ref{tab:results}, both multimodal settings outperformed the unimodal approaches in overall scores, suggesting that combining various modalities provides complementary information that enhances the comprehension of authentic classroom discourse.
Integrating audio signals into the textural context yielded the best average QWK score of 0.384 among all tested modalities, particularly with a notable performance boost in the \textit{Nature of Discourse} component. 
In line with the unimodal findings, adding visual patterns did not lead to further improvement over the text-audio combination.

\begin{table*}[h]
    \caption{Results for performance comparison of different encoders, task configurations, and loss functions (CE: cross-entropy, OLL: ordinal log-loss).}
    \label{tab:comparison}
    \centering
    \renewcommand{\arraystretch}{1.2}
    \begin{tabular}{p{1.5cm}p{2cm}p{1cm}p{2cm}p{2cm}p{2cm}p{2cm}}
        \hline
        \textbf{Encoder} & \textbf{Task} & \textbf{Loss} & \textbf{Nature of} \newline \textbf{Discourse} & \textbf{Questioning} & \textbf{Explanations} & \textbf{Average} \\
        \hline
        LSTM & Multi-task & OLL & 0.512 (0.02) & 0.262 (0.03) & 0.183 (0.05) & 0.319 (0.02) \\
        \hline
        \multirow{4}{*}{Attention}
        & Single-task & OLL & \textbf{0.528} (0.03) & 0.380 (0.04) & \textbf{0.303} (0.06) & \textbf{0.404} (0.02) \\ 
        \cline{2-7}
        & \multirow{3}{*}{Multi-task} & L1 & 0.492 (0.03) & 0.334 (0.02) & 0.252 (0.04) & 0.359 (0.02) \\
        & & CE & 0.503 (0.01) & 0.311 (0.02) & 0.181 (0.07) & 0.332 (0.02) \\
        & & OLL & 0.486 (0.04) & \textbf{0.382} (0.02) & 0.284 (0.04) & 0.384 (0.01) \\
        \hline
    \end{tabular}
\end{table*}

\subsection{Impact of Attention Encoders, Multi-Task Learning, and Ordinal Classification}

We further investigated the impact of key elements in our architecture—namely, the attention-based encoder, multi-task learning, and ordinal loss function—on model performance. In this ablation study, we utilized the best-performing text-audio fusion setting and altered an element each time, keeping other configurations constant for a fair comparison. The results are presented in Table~\ref{tab:comparison}.

\subsubsection{Attention- vs. LSTM-Based Encoder}
LSTM is a recurrent neural network frequently used to model the temporal dynamics of sequential data. Its effectiveness has been demonstrated in educational data analysis \cite{buhler2024lab,ramakrishnan2021toward,buhler2023automated}.
To compare LSTM with our attention-based model, we applied two multi-layer bidirectional LSTMs to encode the textual $\boldsymbol{X}_i^t$ and acoustic $\boldsymbol{X}_i^a$ representations, respectively. 
We then concatenated the last forward and reverse hidden states of the last layer and combined both modalities, yielding a final representation $\boldsymbol{x}_i \in \mathbb{R}^{768 \times 2+1024 \times 2}$ as input to MLPs for prediction. 

Except for \textit{Nature of Discourse}, the attention-based encoder (last row in Table~\ref{tab:comparison}) outperformed the LSTM-based model (first row) across other components and in overall performance.
This improvement can be attributed to two factors: First, the cross-attention encoder potentially captures richer inter-modal interactions. Second, the self-attention mechanism can directly model relationships between any pairwise timestamps and selectively focus on relevant features, making it effective in analyzing complex temporal patterns within classroom discourse. Besides, attention mechanisms process sequential input in parallel, benefiting from computational efficiency.

\subsubsection{Multi- vs. Single-Task Learning} 
In single-task learning, we reported the best-performing result for each component by individually training the model to assess that specific component solely. 
As observed in Table~\ref{tab:comparison}, the component specialists (second row) generally yielded slightly higher or comparable performance across three components, notably achieving a peak QWK score of 0.528 for \textit{Nature of Discourse}. 
This suggests that dedicating separate models to each task enables the encoders to focus on learning distinctive features of each discourse component more precisely.
However, multi-task learning constructs a single model for joint assessment with only a marginal decrease in performance, revealing the efficiency and generalizability of the shared representation that learns the common aspects of classroom discourse. 
This capability can be beneficial specifically in the classroom observation context, where observation protocols typically involve codifying tens of teaching quality dimensions. 

\subsubsection{Regression vs. Standard vs. Ordinal Classification}
We further investigated the advantages of approaching the teaching quality evaluation problem through ordinal classification, compared to regression and standard classification. 
For regression, the model was trained using L1 loss to generate score estimations $y_i^c \in \mathbb{R}$. Since the QWK metric requires categorical input, we rounded these continuous scalars $y_i^c$ to the nearest rating class such that $y_i^c \in \boldsymbol{L}$.
In the case of standard classification, cross-entropy (CE) loss was employed.
To ensure fair comparisons, we assigned class weights to each sample when calculating both losses (like $w^c$ in Equation (\ref{equ:oll})).

As depicted by the overall performance in Table~\ref{tab:comparison}, it appears less effective to tackle the scoring task with standard classifiers. The use of OLL showed additional improvement over L1 loss, particularly in \textit{Questioning} and \textit{Explanations}. 
Unlike CE, which treats all misclassifications as equally wrong, OLL accounts for the inherent order of the discourse ratings by assigning penalties that push predictions closer to the ground truth.
Although regressors also aim to minimize distances from the true values, they are naturally designed to handle continuous labels. A post-processing step is required to convert their outputs into discrete categories, which limits their performance in ordinal classification scenarios \cite{frank2001simple}.

\subsection{Relationships between Discourse Quality Scores and Student Outcomes}

\begin{table}[h]
    \caption{Pearson correlations of human-rated and model-predicted discourse component scores with student outcomes ($N=958$).}
    \label{tab:student_outcomes}
    \centering
    \renewcommand{\arraystretch}{1.2}
    \begin{tabular}{p{2.8cm}P{1.3cm}P{1.3cm}P{1.3cm}}
        \hline
         & \textbf{Test scores} & \textbf{Personal interest} & \textbf{Self-efficacy} \\
        \hline
        \textbf{Nature of Discourse} & & & \\
        \hspace*{0.3cm}Human ratings & -.007 & -.107** & -.021 \\
        \hspace*{0.3cm}T & -.043 & -.100** & -.069* \\ 
        \hspace*{0.3cm}T+A & -.087** & -.098** & -.083* \\ 
        \hspace*{0.3cm}T+A+V & .050 & -.068* & -.053 \\ 
        \hline
        \textbf{Questioning} & & & \\
        \hspace*{0.3cm}Human ratings &  .059 & -.036 & .034 \\
        \hspace*{0.3cm}T & .031 & -.116*** & -.058 \\ 
        \hspace*{0.3cm}T+A & .001 & -.162*** & -.086** \\
        \hspace*{0.3cm}T+A+V & -.043 & -.108** & -.095** \\
        \hline
        \textbf{Explanations} & & & \\
        \hspace*{0.3cm}Human ratings & -.058 & .089** & .064 \\
        \hspace*{0.3cm}T & .099** & .011 & .069* \\ 
        \hspace*{0.3cm}T+A & .083* & .064 & .069* \\
        \hspace*{0.3cm}T+A+V & -.026 & .014 & .062 \\ 
        \hline
        \multicolumn{4}{p{8cm}}{\textit{Notes.} * $p < 0.05$, ** $p < 0.01$, *** $p < 0.001$}
    \end{tabular}
\end{table}

Ultimately, we examined the associations between human-rated and model-predicted discourse quality scores and student learning and non-cognitive outcomes. Following the original GTI study \cite{gti2020}, we computed classroom-level discourse component scores by averaging the ratings across segments and then lessons for the same teacher.
Three student outcome indicators, collected after instruction, were used: test scores measuring students' knowledge of quadratic equations, along with self-reported interest and self-efficacy in mathematics. 
We tested estimations from models using different modalities. The Pearson correlation results are shown in Table~\ref{tab:student_outcomes}. 
Regarding \textit{Explanations}, human ratings showed a statistically significantly positive correlation with student interest, while two models' predictions (text-only and text-audio) were positively related to student test achievements and self-efficacy. 
The statistically significant negative associations of predictions for \textit{Nature of Discourse} and \textit{Questioning} with student interest aligned with findings of the GTI Germany study where the quality of instruction (discourse is one of the four sub-domains included by the general instruction domain) also reported a negative correlation with student interest \cite{talisgermany2020}.
Overall, although we generally found weak associations between discourse quality scores and student outcomes, the correlation analysis suggested the promise of our automated methods as a valuable proxy to replicate those findings based on human ratings in large-scale classroom observation studies.

\section{Discussion}

\subsection{Main Findings}
Leveraging human IRR as a benchmark, we demonstrated the efficacy of our proposed architecture in assessing discursive practices across various dimensions. 
In terms of overall performance, even the exclusive use of textural representations achieved a slightly higher agreement with the manual ratings than human IRR.
The text-guided integration of acoustic features further increased the model's predictive accuracy by capturing nuances in speech that text alone could not convey, surpassing human IRR by a notable margin. 
By comparing different modalities, we found that textual representations provided the most valuable insights into discourse assessment, particularly for components requiring semantic understanding such as question quality and explanation depth identification. 
This advantage likely stems from the fact that transcripts record the content of classroom dialogue, facilitating the analysis of discourse substance and structure.
In addition, the effectiveness of the textual pathway was enhanced by the adopted pre-trained textual encoder's ability to handle longer input sequences, allowing it to consider the context within an entire segment (Sect.~\ref{subsubsec:text_encoder}).
Moreover, acoustic representations made a specific contribution to estimating \textit{Nature of Discourse}, revealing their capability to offer complementary information.
Remarkably, we found that the inclusion of visual features did not improve the prediction accuracy. 
Similar findings were observed in a recent study on multimodal sentiment analysis \cite{wu2024multimodal}. 
This could be attributed to both the limited relevance of visual cues in discourse assessment and the challenge of extracting meaningful patterns from authentic classroom videos due to factors such as frequent occlusions and the complexity of classroom interactions.
Our ablation study further showed the effectiveness of attention-based encoders in capturing inter- and intra-modal interactions, the trade-off provided by multi-task learning between performance and computational efficiency, and the benefits of formulating automated assessment as an ordinal classification task. 
The correlation analysis between predicted discourse quality scores and student outcomes revealed partially similar patterns to those observed using human ratings, indicating that developing automated tools could facilitate future educational research into classroom observation instruments' quality, validity, and reliability.

\subsection{Implications and Ethical Considerations}
Our findings underscore the potential of the proposed approach to automate the resource-intensive classroom observation process, thus promoting research on teaching effectiveness assessment at scale.
Prior research has shown that these automated assessments could provide teachers with scalable and frequent feedback on multidimensional constructs of teaching, encouraging reflection and improvement in their instructional practices \cite{whitehill2023automated,wang2023chatgpt}.
Specifically, by accurately assessing question quality and explanation depth, the model could help teachers refine their communication strategies and leverage intentional discourse to foster student critical thinking and learning.
Our study offers methodological insights into applying novel multimodal fusion techniques to obtain more robust and accurate predictions for the holistic assessment of teacher-student interactions. 

Privacy considerations are critical, especially in classroom environments involving minors. 
A significant finding of our study is that relying solely on audio data (which can be transcribed into text) can maintain high performance for classroom discourse assessment, eliminating the need for video recordings that could expose sensitive facial information. This practical approach would not only streamline the data collection process but also offer a more privacy-preserving solution.
Moreover, the use of frozen unimodal encoders also fosters privacy protections. On one hand, it allows for the immediate deletion of raw recordings after feature extraction \cite{sumer2021multimodal}. On the other hand, unlike hand-crafted representations, the extracted latent embeddings are entirely uninterpretable, which prevents the inference of original content.

Beyond privacy, implementing automated discourse assessment systems should adhere to key ethical principles to ensure responsible and equitable use. First, the collection and use of classroom data must be transparent and consensual. All participants, including teachers and students, should be fully informed of what data is being collected, how it will be used, and their right to opt-out. 
Moreover, fairness and inclusiveness are critical for AI-driven educational applications. Biases in model performance, particularly across different student demographics, cultural contexts, or teaching styles, could lead to unfair evaluations \cite{nguyen2023ethical}. To mitigate this risk, it is essential to ensure that training data are diverse and representative of varied classroom settings and teaching practices.
Additionally, we emphasize such automated assessments should be used as supportive tools for teacher self-reflection and professional growth. Teachers should maintain agency in using such voluntary, self-directed feedback to improve their instructional practices rather than being subject to automated judgments.

\subsection{Limitations and Future Work}
Despite the promising findings, this study has several limitations requiring consideration in future research. 
First, although the assessment of classroom discourse appears to rely less on visual cues, the limited performance of the visual pathway may also stem from the ineffectiveness of the extracted CLIP representations in our settings. Future studies could explore more advanced computer vision techniques or integrate individual features such as gesture and facial expression analysis. Likewise, different pre-trained models could be evaluated to process transcripts and audio signals. 
Second, generalizability is always a key concern in the field of machine learning. Although our evaluation method considers generalizability to new teachers, the used dataset typically does not cover the full diversity of classroom settings. Given the international GTI study, future research could involve datasets beyond Germany and carry out cross-cultural experiments to evaluate how the model performs and how the assessments relate to student outcomes in diverse educational systems.
Another limitation lies in model explainability. Applying Explainable AI (XAI) techniques would be a valuable future direction. For instance, visualizing attention heatmaps \cite{yang2018ncrf} could highlight those classroom utterances or keywords that contribute significantly to model inferences. This transparency could enhance human trust in AI and potentially offer educators valuable insights to refine pedagogical strategies.
Besides, although our current study focuses on classroom discourse, the proposed architecture is extensible to other facets of teaching effectiveness. Thus, one future objective is to adapt and evaluate the method for other dimensions like \textit{Student Cognitive Engagement} in the GTI observation protocol.
Finally, some correlation results, such as the negative relationship between student-centered discourse and interest (noted in the original GTI report), seem counter-intuitive. These findings may be influenced by how the coding rubrics were constructed, which partially reflected the difficulty and complexity of instruction. Thus, integrating pedagogical expertise through a human-in-the-loop approach would be essential to interpret these patterns and inform decision-making.
We also intend to strengthen collaborations with educators to gather their perceptions of the plausibility of the predicted scores, promoting the model's applicability.

\section{Conclusion}
The present study introduces an attention-based multimodal fusion architecture that utilizes multi-task learning to simultaneously assess three classroom discourse dimensions (i.e., \textit{Nature of Discourse}, \textit{Questioning}, and \textit{Explanations}), yielding a level of agreement with human raters comparable to inter-rater reliability in manual coding.
The superior performance achieved by integrating textual and acoustic features underscores not only the importance of exploiting multimodal data but also the practicality of employing only audio recordings.
This effective and efficient approach suggests a pathway toward large-scale classroom observation research, potentially offering teachers timely and valuable feedback on the quality of their teaching practices.


%
\bibliographystyle{abbrv}
\bibliography{mybibliography}  
\end{document}